\documentclass[final,3p,times,twocolumn]{elsarticle}
 \biboptions{comma,sort&compress}
\usepackage{here}
\usepackage{graphicx}
\usepackage{epsfig}
\usepackage{hyperref}
\usepackage{rotating}
\def\nin{\noindent}
\def\beq{\begin{equation}}
\def\eeq{\end{equation}}
\def\bea{\begin{eqnarray}}
\def\eea{\end{eqnarray}}

\journal{Nuc. Phys. (Proc. Suppl.)}

\begin{document}

\begin{frontmatter}



\title{A {\textit{Mathematica}} interface to NNPDFs}

 \author[label1]{Nathan P. Hartland}
  \address[label1]{Tait Institute, School of Physics and Astronomy, University of Edinburgh, EH9 3JZ, UK.}
   \author[label2]{Emanuele R. Nocera}
  \address[label2]{Dipartimento di Fisica, Universit\`{a} di Milano and INFN, Sezione di Milano, Via Celoria 16 I-20133 Milano, Italy}
%\author{}

%\address{}

\begin{abstract}
\noindent
We present a {\textit{Mathematica}} interface for handling the parton distribution functions of the NNDPF Collaboration,
available from the NNPDF hepforge website~\url{http://nnpdf.hepforge.org/}.
As a case study we briefly summarise the first PDF set which includes all relevant LHC data, NNPDF2.3, and
demonstrate the use of our new {\textit{Mathematica}} interface.

% We present a new interface to the NNPDF parton
% distributions in form of a \textit{Mathematica} package, available from the
% NNPDF hepforge website~\url{http://nnpdf.hepforge.org/}. We illustrate the main
% features of this interface, in particular the plotting and the statistical modules. 
% We demonstrate these capabilities with
% the recent NNPDF2.3 parton set, the first PDF set that includes
% all relevant LHC data for PDF constraints.
%We present a new \textit{{Mathematica}} package for handling the parton distribution functions of the NNPDF Collaboration. As a case study we briefly summarise the NNPDF2.3 parton set and demonstrate the use of the new \textit{{Mathematica}} package. 
\end{abstract}


\end{frontmatter}

% \linenumbers

%\section{Introduction}
%\label{}
\nin
%%%%%%%%%%%%

We have developed a \textit{Mathematica} 
package which provides direct interactive access to any NNPDF parton set.
This package reads the  NNPDF LHgrid files as available from the LHAPDF library
and performs their interpolation in ($x$,$Q^2$) space by means of proper 
{\textit{Mathematica}} built-in functions.   
The user can therefore make use of PDF ensembles within a {\textit{Mathematica}}
notebook in order to compute observables, evaluate PDF central values and variances and plot
PDFs instantaneously. 
In this note we shall present a brief motivation and summary of the key features present 
in the new NNPDF2.3~\cite{Ball:2012cx} PDF set, 
and demonstrate the NNPDF \textit{Mathematica} package 
by examining this new determination.

%\section{NNPDF2.3}
In recent years the volume of collider data of application in the accurate determination of 
parton distribution functions has significantly increased. 
Previous PDF determinations based solely upon high energy collider data collected by 
Tevatron and HERA collaborations were poorly constrained in comparison to global fits including
data from fixed target deep inelastic scattering and Drell-Yan experiments. 
These collider only determinations are of particular interest in that they are free from the potential
issues arising from nuclear corrections and higher twist effects present in the low energy data of 
global fits. 
Studying the impact of LHC data upon PDF determinations is therefore crucial not only for further constraint 
upon global determinations, but also for their potential utility in providing competitive collider only fits.

The NNPDF2.2 parton set~\cite{Ball:2011gg} was the first determination to investigate 
the constraining power of LHC measurements upon parton distributions, specifically the $W$ lepton charge asymmetry 
measurements of the ATLAS~\cite{Aad:2011yna} and CMS~\cite{Chatrchyan:2011jz} collaborations in addition to the 
D0 collaboration measurements~\cite{Abazov:2008qv,Abazov:2007pm}. 
This data was included by taking advantage of the Monte Carlo uncertainty estimation 
in the NNPDF methodology. NNPDF sets provide a probability distribution in the space of 
PDFs allowing for the use of a Bayesian reweighting technique \cite{Ball:2010gb,Watt:2012tq}
 and the rapid inclusion of new datasets into a prior set. 
However, this method is limited to the inclusion of a relatively small dataset.

In order to include an updated and more comprehensive LHC dataset into an NNPDF fit, 
an extremely fast method of collider observable computation has been developed.
The FastKernel method for hadronic observables~\cite{Ball:2012cx} combines the Monte Carlo weight grids 
from fast NLO software packages such as APPLgrid~\cite{arXiv:0911.2985} 
and FastNLO~\cite{Kluge:2006xs}, with FastKernel PDF evolution tables~\cite{Ball:2010de}. 
The resulting FastKernel table may then be combined with the initial scale parton distributions
to produce the required observable. For a given hadronic observable $\sigma$ the operation 
required to compute the theoretical prediction for such an observable is simply,
\begin{equation}
\label{eq:FKsigma1} 
  \sigma_I = \sum_{i,j}^{N_{\mathrm{pdf}}} \sum_{\alpha,\beta}^{N_x} 
  \Sigma^I_{\alpha\beta i j} N_{\alpha i}^0N_{\beta j}^0
  \mbox{ ,}
  \end{equation}
where $\Sigma$ is the final FastKernel table and the $N$ are the initial scale, 
DGLAP evolution basis parton distributions. The indices $i$, $j$ run over the $N_x$ 
points in the fitting scale $x$-grid, and $\alpha$, $\beta$ run over the reduced flavour basis 
of $N_{\mathrm{pdf}}$ light PDFs at this initial scale. 
This operation is computationally very simple and allows for a much accelerated fit.

\begin{figure}[ht]
    \begin{center}
      \includegraphics[width=0.48\textwidth]{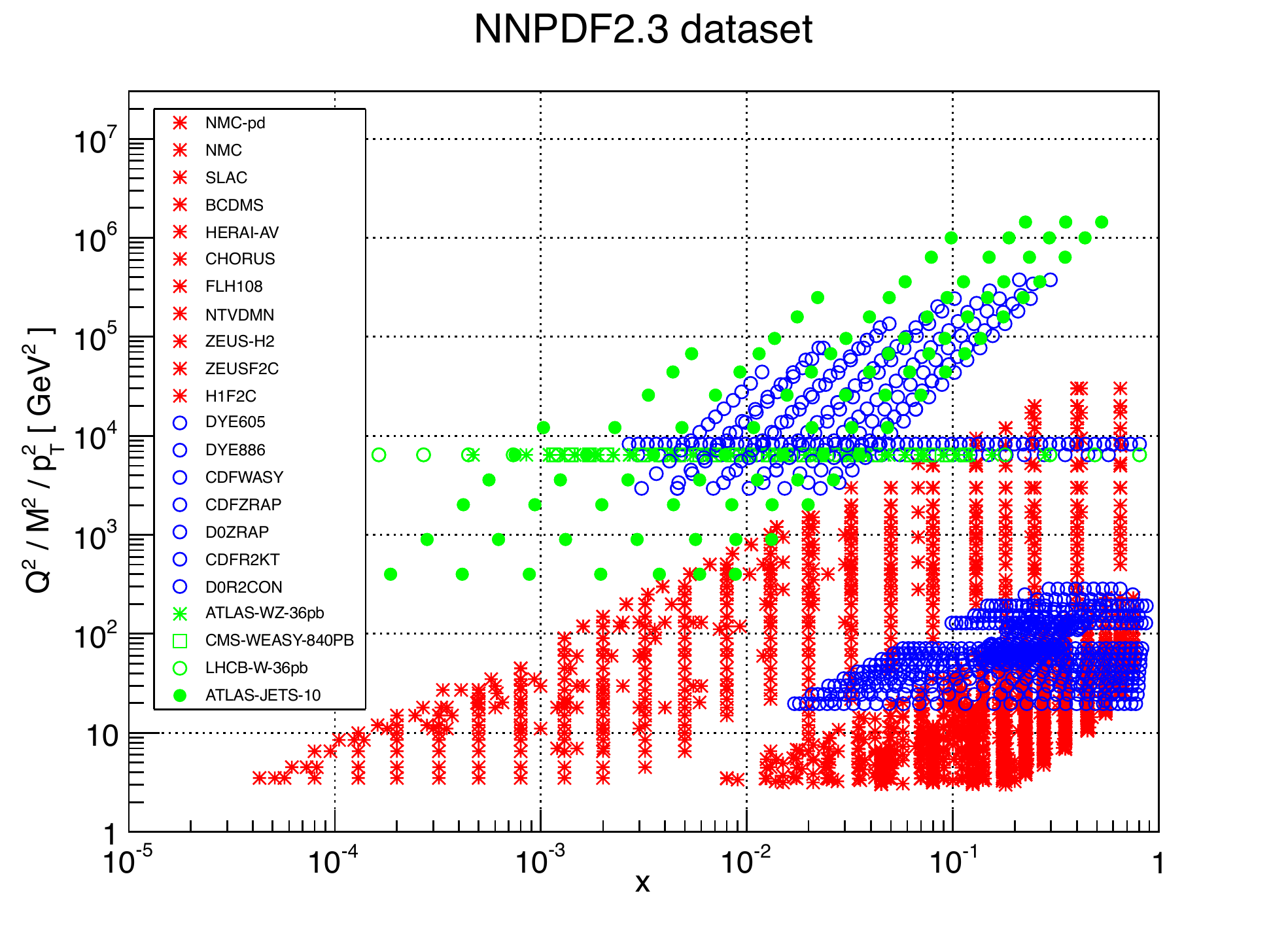}
    \end{center}
    \vskip-0.5cm
    \caption{\small Kinematic range of the NNPDF2.3 dataset.}
    \label{fig:kin}
\end{figure}

With this method we are able to add a relatively large LHC dataset to the existing data in the 
NNPDF 2.1 family to produce a set of fits denoted NNPDF 2.3. 
The NNPDF2.3 LHC dataset includes ATLAS collaboration $35$ $\mathrm{pb}^{-1}$ measurements of 
the inclusive jet cross section~\cite{Aad:2010ad} and electroweak vector boson rapidity 
distributions~\cite{Aad:2011dm}. 
The CMS collaboration's $840$ $\mathrm{pb}^{-1}$ measurement of the W electron asymmetry~\cite{CMSW840} 
and the LHCb W boson rapidity distribution data~\cite{Aaij:2012vn} are included also. 
The expanded kinematic reach of the NNPDF2.3 dataset is described in Fig.~\ref{fig:kin}. 
In addition to the inclusion of the LHC data, the NNPDF2.3 family of fits features a number 
of methodological improvements. 

\begin{figure}[p]
    \begin{center}
      \includegraphics[width=0.4\textwidth]{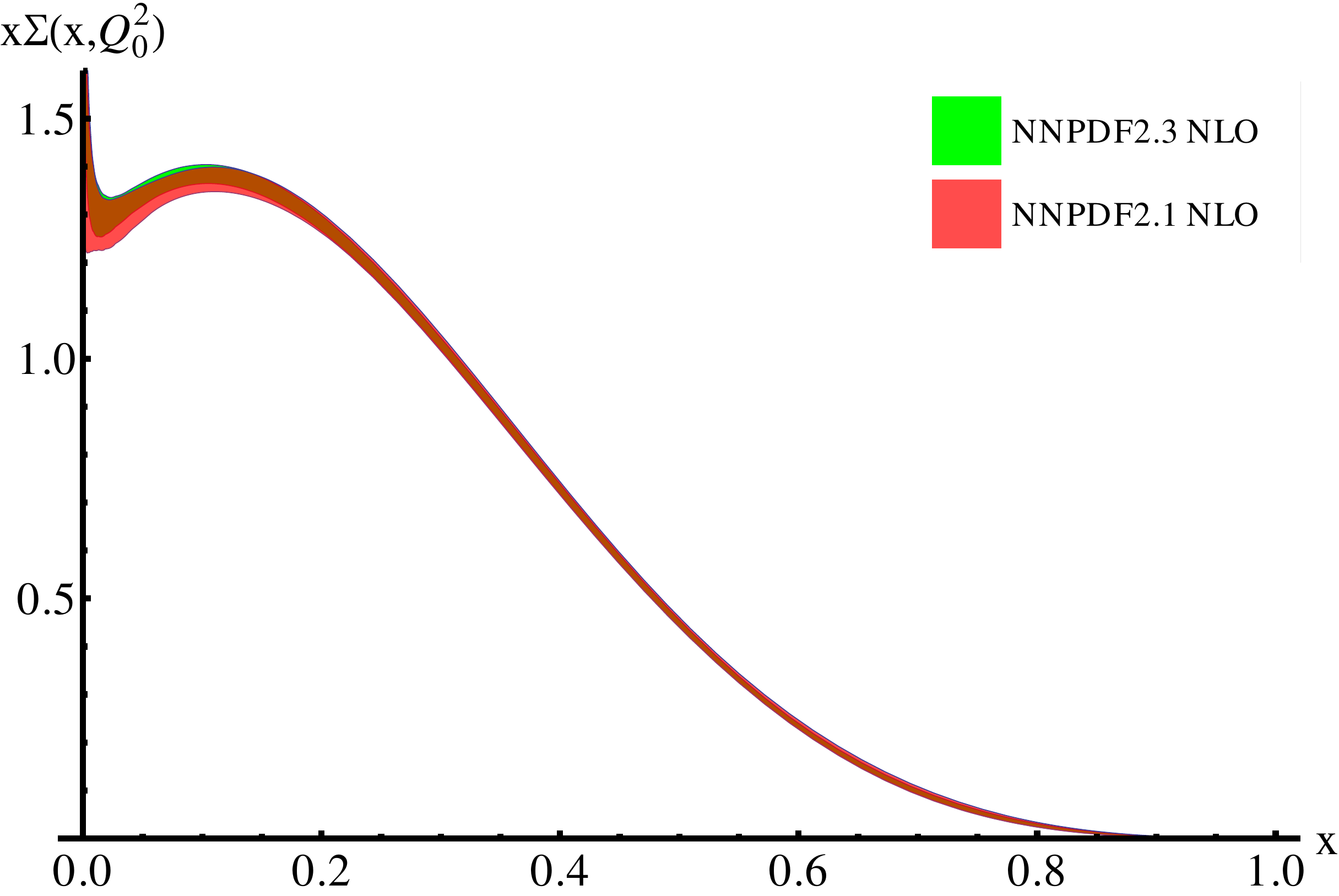} \\
      \ \\ \ \\
      \includegraphics[width=0.4\textwidth]{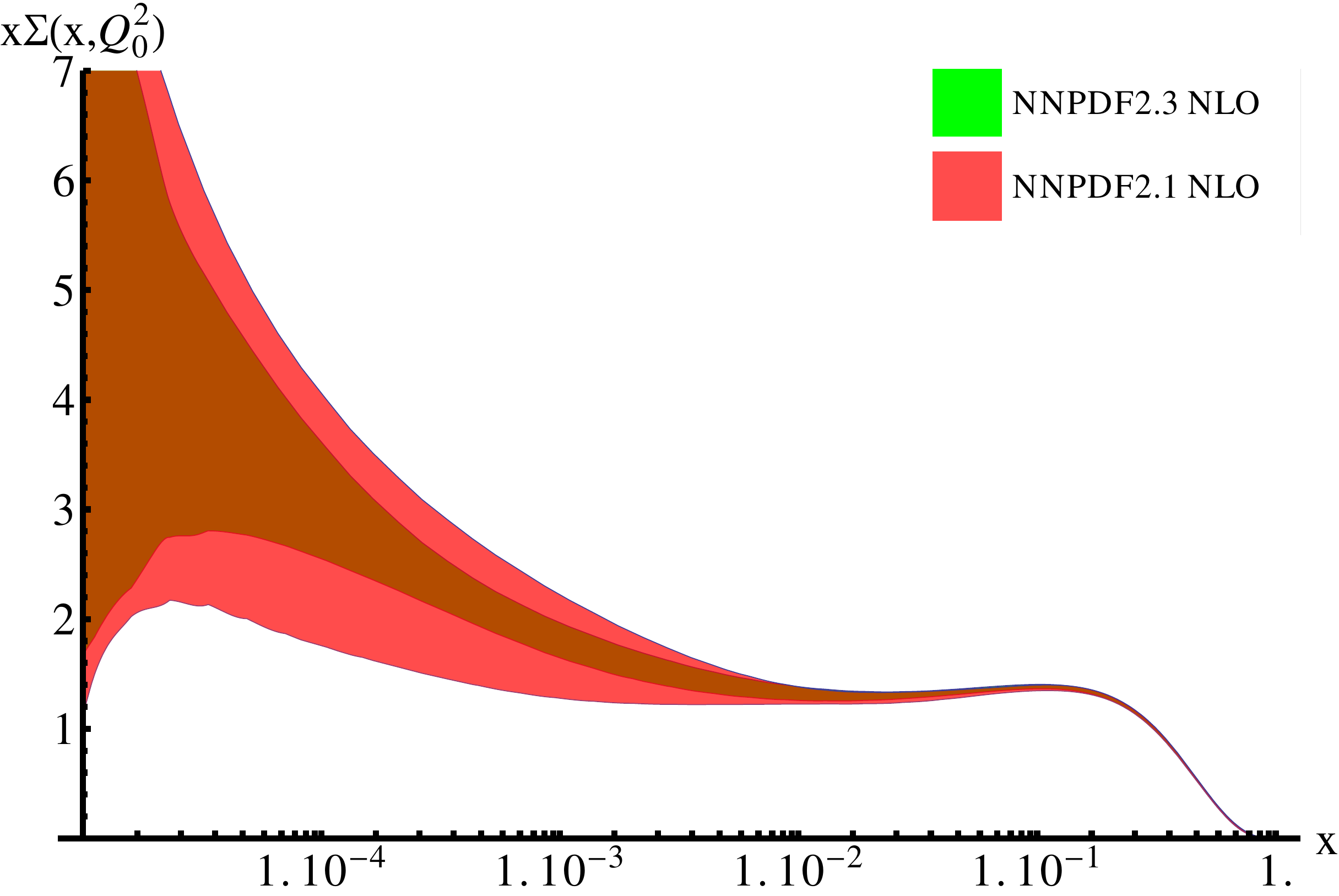} \\ 
      \ \\ \ \\
      \includegraphics[width=0.4\textwidth]{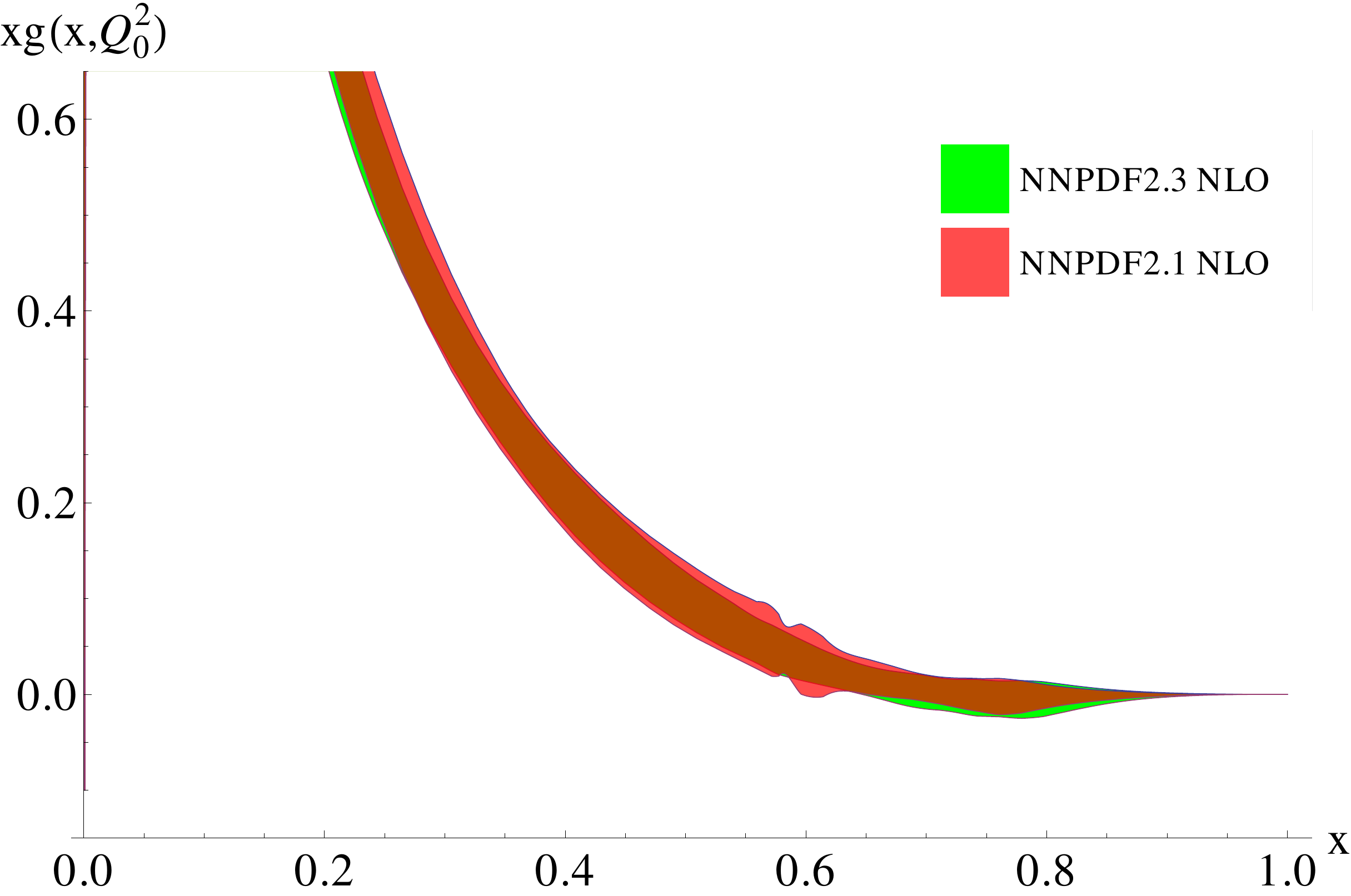} \\
      \ \\ \ \\
      \includegraphics[width=0.4\textwidth]{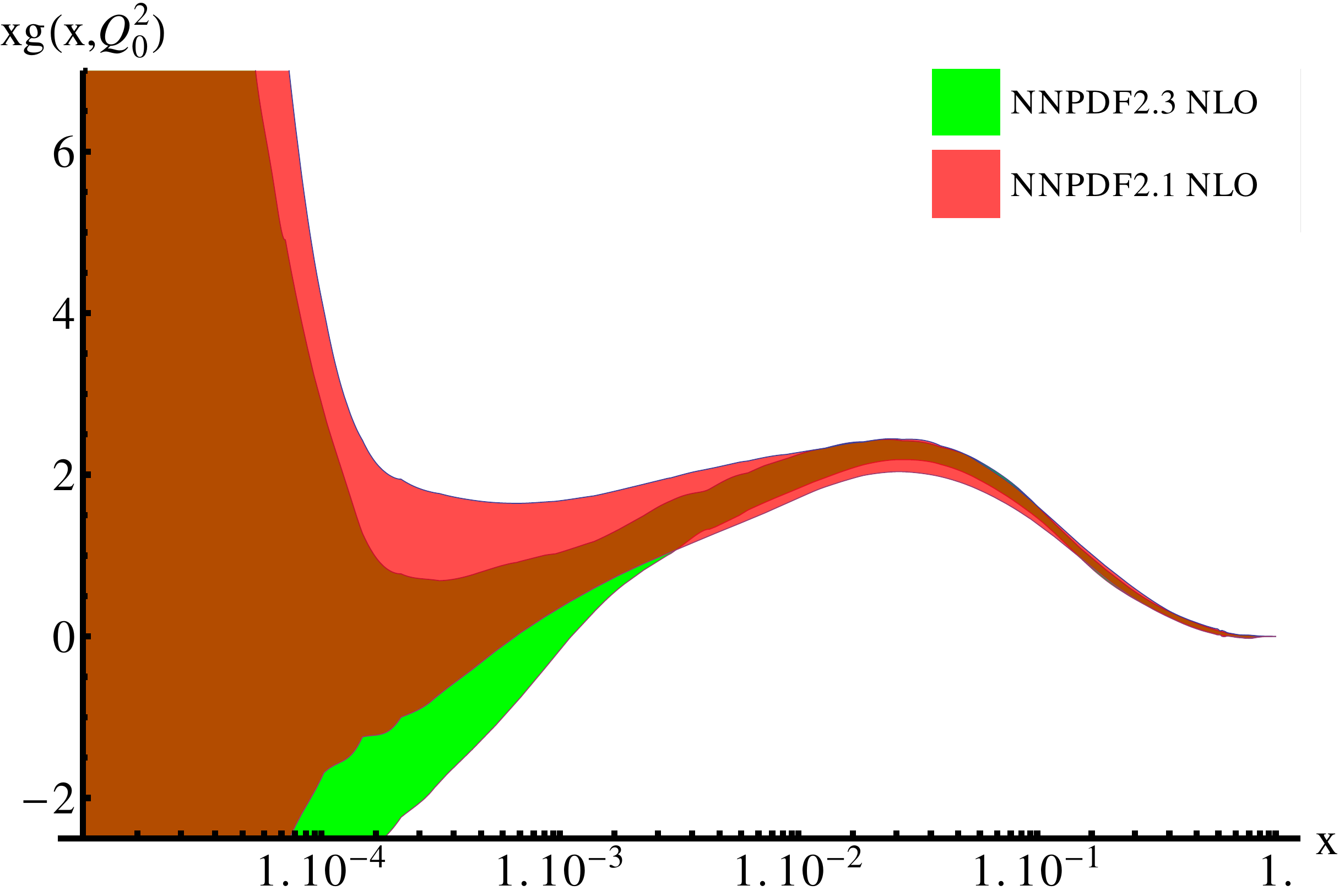} \\
    \end{center}
    \vskip-0.5cm
    \caption{\small NLO NNPDF2.3 (green) singlet sector PDFs at $Q_0^2 = 2$ GeV$^2$, compared to their NNPDF2.1 (red)
counterparts. All error bands shown correspond to a one sigma interval. These plots are drawn with 
the new NNPDF {\textit{Mathematica}} interface.}
    \label{fig:plotcomp}
\end{figure}

In order to assess the impact of the included LHC data, we compare the 
NLO NNPDF2.3~\cite{Ball:2012cx} and NLO NNPDF2.1~\cite{Ball:2011mu} 
singlet sector PDFs at $Q_0^2=2$ GeV$^2$ (see Fig~\ref{fig:plotcomp}).
We have used our new {\textit{Mathematica}} interface to draw these plots.
From Fig.~\ref{fig:plotcomp}, it is clear that PDFs
from the two sets differ by less, and usually much less, than one sigma.

%\section{{\textit{Mathematica}} interface availability}

\begin{figure*}[p]
    \begin{center}
      \includegraphics[scale=0.43,angle=90]{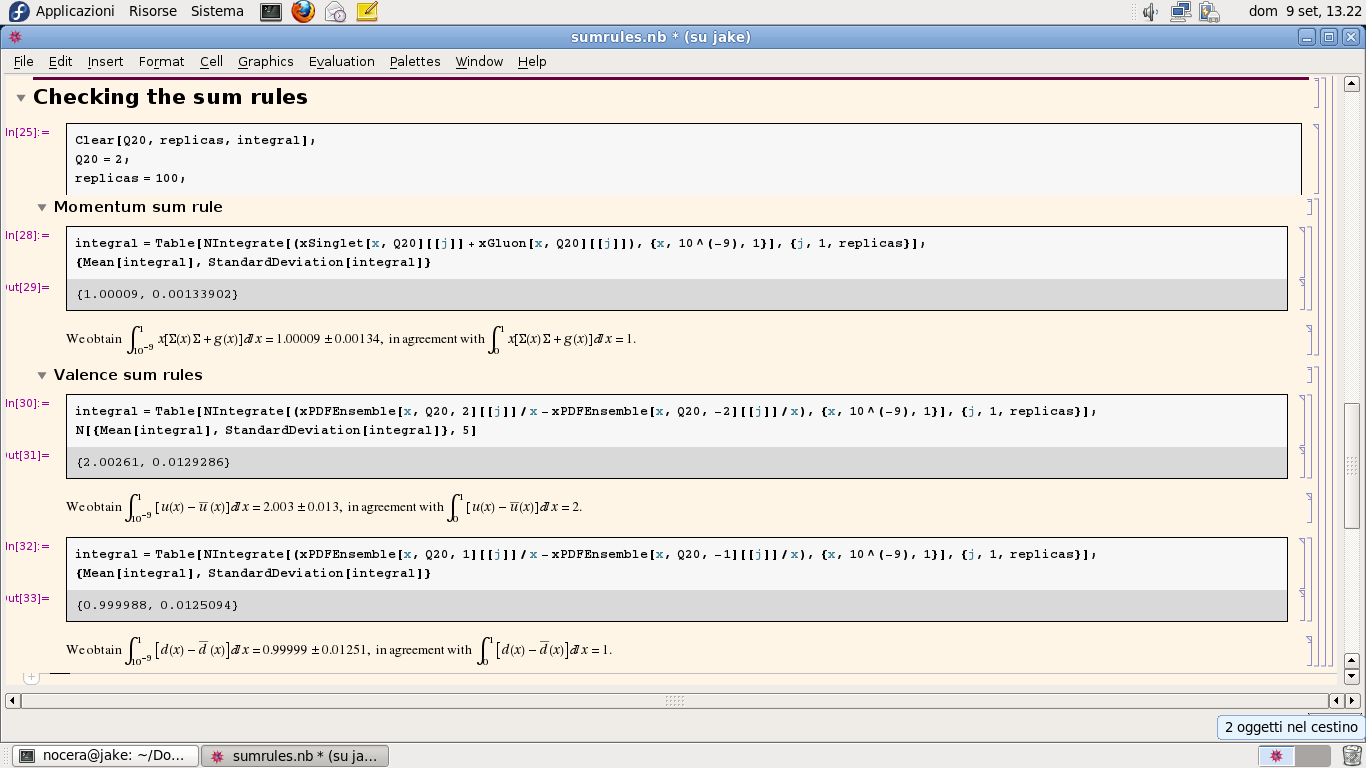}
    \end{center}
    \vskip-0.5cm
    \caption{\small A snapshot of the {\textit{Mathematica}} notebook
written with our interface for checking the momentum and valence sum rules from the NLO NNPDF2.3 parton set.}
    \label{fig:snapshot}
\end{figure*}

The main advantages of our interface come from the possibility to combine the use of any NNPDF set
with all the \textit{Mathematica} features. In particular, this allows to perform a variety of manipulations on PDFs
in a straightforward and interactive way, as we briefly demonstrate in the following discussion.
\begin{itemize}
 \item {\bf{Compute PDF central value and variance}} We have defined proper functions to keep the computation 
of PDF central value and variance very easy. These built-in functions only need $x$, $Q^2$ and PDF flavour as input.
The user can also specify the confidence level to which central value and variance should be computed.
 \item {\bf{Make PDF plots}} \textit{Mathematica} enables a wide range of plotting options. As a few examples, we
show the 3D plot and the contour plot of the Singlet PDF from 
the NLO NNPDF2.3 parton set (see Figs.~\ref{fig:xSinglet3D} - \ref{fig:xSingletCP} respectively). 
\begin{figure}[t]
    \begin{center}
     \includegraphics[width=0.4\textwidth]{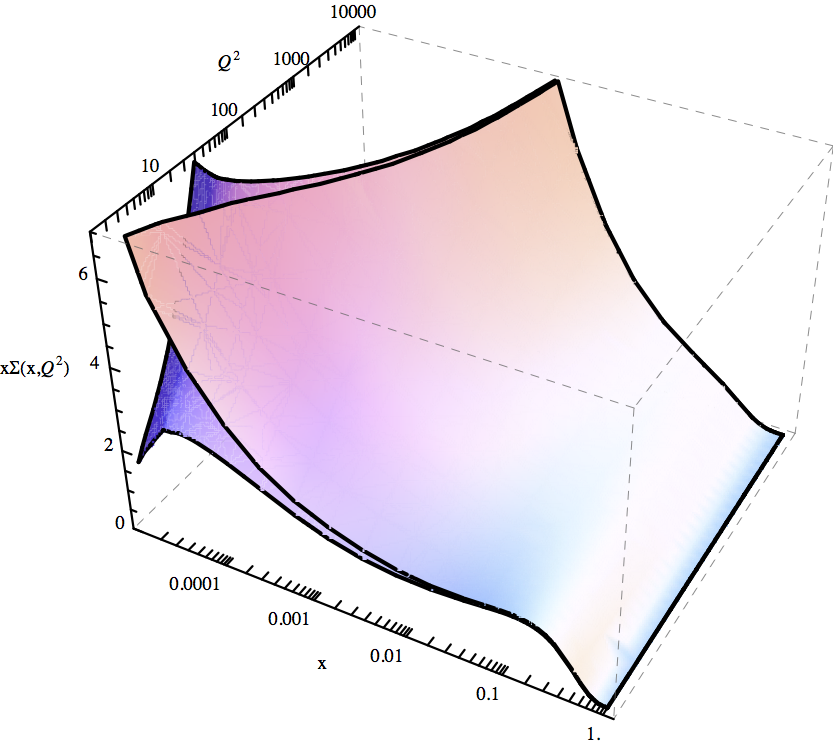}
    \end{center}
    \vskip-0.5cm
    \caption{\small Simultaneous dependence from both $x$ and $Q^2$ for the one sigma error of the
NLO NNPDF2.3 Singlet PDF. This plot is drawn with 
the new NNPDF {\textit{Mathematica}} interface.}
    \label{fig:xSinglet3D}
\end{figure}
\begin{figure}[t]
    \begin{center}
      \includegraphics[width=0.4\textwidth]{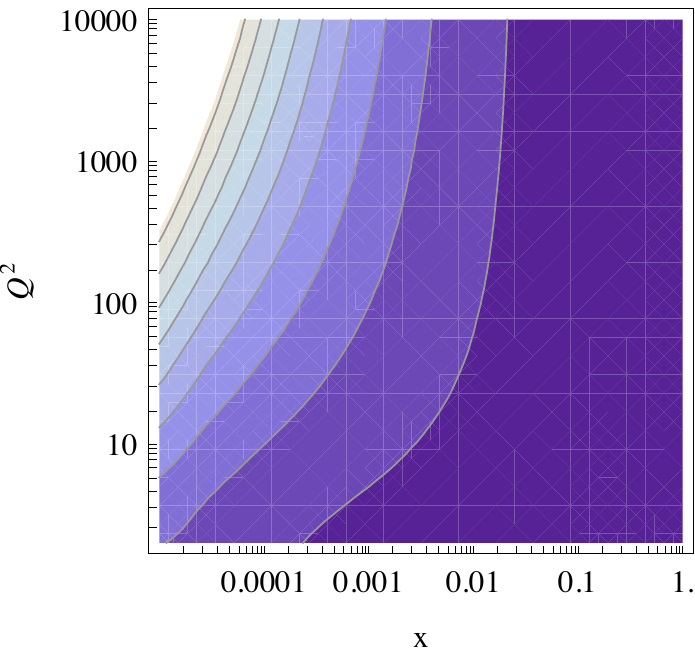}
    \end{center}
    \vskip-0.5cm
    \caption{\small Contour plot of the NLO NNPDF2.3 Singlet PDF in the ($x$,$Q^2$) plane. 
This plot is drawn with the new NNPDF {\textit{Mathematica}} interface.}
    \label{fig:xSingletCP}
\end{figure}
\item {\bf{Perform computations involving PDFs}} PDF manipulation can be carried out straighforwardly
since we have defined 
functions which handle either single replicas or the whole Monte Carlo ensemble. 
The user can then easily perform any computation which involves PDFs.
For example, we show in Fig.~\ref{fig:snapshot} a snapshot 
of a typical {\textit{Mathematica}} notebook in which we use our interface to
check the momentum and valence sum rules from the NLO NNPDF2.3 parton set.
\end{itemize}

The new NNPDF {\textit{Mathematica}} package can be downloaded 
from the NNPDF web site,
\begin{center}
{\bf \url{http://nnpdf.hepforge.org/}}
\end{center}
together with a sample notebook containing a step by step explanation of the NNPDF usage within
{\textit{Mathematica}} as well as a variety of examples.

\section*{Acknowledgments}
The authors would like to thank the organisers of QCD 2012 for the opportunity to present this work. 
This contribution made use of resources provided by the Edinburgh Compute and Data Facility (ECDF) 
(\url{http://www.ecdf.ed.ac.uk/}). The ECDF is partially supported by the eDIKT initiative 
(\url{http://www.edikt.org.uk}).
%\section*{References}

\end{document}